\documentstyle[pre,preprint,aps]{revtex}
\tightenlines
\newcommand{\siml}{\stackrel{<}{\sim}}
\newcommand{\simg}{\stackrel{>}{\sim}}

\begin{document}
\draft
%\baselineskip=0.5\baselineskip
\title{
Responses of Ensemble Neurons 
to Spike-Train Signals \\
with Independent Noises:
Stochastic Resonance \\
and Spike Variability
%Spike-Train and/or White Noises 
%subject to Spike-Train and White Noises
%\footnote{E-print: cond-mat/0111xxx}
}
\author{
Hideo Hasegawa
\footnote{E-mail:  hasegawa@u-gakugei.ac.jp}
}
\address{
Department of Physics, Tokyo Gakugei University,
Koganei, Tokyo 184-8501, Japan
}
\date{\today}
\maketitle
\begin{abstract}
Responses have been numerically
studied of an ensemble of $N$ (=1, 10, and 100) 
Hodgkin-Huxley (HH) neurons 
to coherent spike-train inputs applied
with independent Poisson spike-train (ST) noise and
Gaussian white noise.
Three interrelated issues have been investigated:
(1) the difference and the similarity between the effects of
the two noises,
(2) the size effect of a neuron ensemble on 
the signal-to-noise ratio (SNR), and
(3) the compatibility of a large firing variability 
with fairly good information transmission.
(1) The property of stochastic resonance (SR) 
for ST noise is shown to 
be rather different
from that for white noise.
When SNR for sub-threshold inputs obtained in our simulation 
is analyzed by the expression given by
$SNR=10\; {\rm log_{10}}\;[(A/X^{\alpha})\;{\rm exp}(-B/X)]$ 
where $X$ expresses the
noise intensity and $A$ and $B$ 
are constants,
the index $\alpha$ is $\alpha=3$  for the ST noise 
and $\alpha=2$ for the white noise: 
the former is different from the conventional
value of $\alpha=2$ realized in many
non-linear systems.
ST noise works less effectively for SR 
than white noise.
(2) The transmission fidelity evaluated by SNR
is much improved by increasing $N$,
the size of ensemble neurons.
In a large-scale neuron ensemble, 
SNR for supra-threshold inputs 
is shown to be not significantly degraded by weak noises
responsible to SR for sub-threshold inputs.
(3) Interspike intervals (ISIs) of output spikes
for sub-threshold inputs 
have a large variability ($c_{v} \siml 0.8$),
which is comparable to the data observed in cortical neurons.
Despite variable firings of individual neurons,
output signals summed over an ensemble may carry
information with a fairly good SNR by the aid of SR
and a pooling effect.

\end{abstract}

\noindent
\vspace{0.5cm}
\pacs{PACS No. 87.10.+e 84.35.+i 05.45.-a 07.05.Mh }
%
%\narrowtext
\section{INTRODUCTION}

%\begin{center}
%{\bf I. INTRODUCTION}
%\end{center}

It has been controversial how neurons communicate 
information by spikes \cite{Rieke96}-\cite{Pouget00}.
Much of debates on the nature of the neural code has been
mainly focused on the two issues.
The first issue is whether information is encoded
in the average firing rate of neurons ({\it rate code})
or in the precise firing times ({\it temporal code}).
%It has been widely believed that information is encoded
%in the average firing rate of individual neurons ({\it rate code}).
Andrian \cite{Andrian26} first noted the relationship
between neural firing rate and stimulus intensity, which
forms the basis of the rate code.
Actually firing activities of motor and sensory neurons are
reported to vary in response to applied stimuli.
In recent years, however, the alternative 
temporal code has been proposed in which detailed
spike timings are assume to play an important role in information
transmission:
information is encoded in interspike
intervals (ISIs) or in relative timings between
firing times of spikes \cite{Softky93}-\cite{Stevens98}.
Indeed, experimental evidences  have accumulated 
in the last several years, indicating a  use of
the temporal coding in neural systems \cite{Carr86}-\cite{Thorpe96}.
Human visual systems, for example, have shown to classify
patterns within 250 ms despite the fact that at least 
ten synaptic stages are involved from retina to 
the temporal brain \cite{Thorpe96}.
The transmission
times between
two successive stages of synaptic 
transmission are suggested to be no more than 10 ms 
on the average.
This period is too short to allow rates to be determined
accurately.

The second issue is whether
information is encoded in the activity of single (or very few) neurons
or that of a large number of neurons
({\it population} or {\it ensemble code}).
The population rate code model assumes that information is
coded in the relative firing rates of ensemble neurons,
and has been adopted in the most of the theoretical analysis
\cite{Abbott98}.
On the contrary, in the population temporal code model,
it is assumed that
relative timings between spikes in ensemble neurons
may be used as an encoding mechanism for perceptional
processing \cite{Hopfield95}-\cite{Rullen01}.
A number of experimental data supporting this code have been reported
in recent years \cite{Gray89}-\cite{Hatso98}.
For example, data have demonstrated that temporally
coordinated spikes can systematically signal sensory
object feature, even in the absence of changes
in firing rate of the spikes \cite{deCharms96}.

The strong criticism against the temporal code is that
spikes are vulnerable to noise while the rate code
performs robustly in the presence of noise but with
limited information capacity.
It is well known that although firings of single neocortical 
neurons in {\it vitro} are precise and
reliable, those in {\it vivo} are quite unreliable \cite{Mainen95}.
This is due to noisy environment in {\it viro},
which makes the reliability of 
neurons firings worse.
In recent years, however, much studies have been made
for the stochastic resonance (SR) \cite{Gammai98}\cite{Anish99}
in which information transmission of signals is enhanced by 
background noises,
against our conventional wisdom.
SR in a neural system 
has been theoretically investigated 
\cite{Bulsara96}-\cite{Liu00}.
The transmission fidelity for weak external signals, which is
evaluated by the signal-to-noise ratio (SNR) or
the peak height of the interspike-interval (ISI) distribution,
is enhanced by added noises.
SR is supported in
some physiological experiments for biological 
systems such as crayfish \cite{Douglass93}\cite{Pei96b},
cricket \cite{Levins96}
and rat \cite{Gluckman96}\cite{Nozaki99}.

Although SR itself is a genetic phenomenon,
its detailed character is determined by
the three important factors: (a) kinds of systems (neurons),
(b) input signal and (c) noises. 
As for the first factor of (a) neurons,
SR in single neurons has been studied by 
using various theoretical models such as
the integrate-and-fire (IF) 
model \cite{Bulsara96}-\cite{Shimokawa99a},
the FitzHough-Nagumo (FN) model \cite{Longtin93}-\cite{Longtin94}
and the Hodgkin-Huxley (HH) model \cite{Lee98}\cite{Lee99}.
SR in coupled or ensemble neurons has been also investigated
by using the IF model \cite{Shimokawa99b}-\cite{Lindner01}, 
FN model \cite{Collins95a}-\cite{Stocks01} 
and HH model \cite{Pei96a}-\cite{Liu00}.
The transmission fidelity has maximum when the noise magnitude
or the coupling strength is changed. 
It has been pointed out that the transmission fidelity 
of ensemble neurons for sinusoidal inputs with
independent white noises is improved
as the size of an ensemble is increased \cite{Collins95a}.

As for the second factor of (b) input signals,
most of theoretical studies have been made
for analog inputs with periodic (mostly sinusoidal) 
or aperiodic amplitude modulation \cite{Collins95b}.
This is because these studies have been motivated by a fact
that peripheral sensory neurons play a role of transducers
receiving analog stimulus and emitting spikes.
In central neural systems, however, cortical neurons 
are reported to play a role of data-processors
receiving and transmitting spike trains \cite{Traub99}.
There are only a few theoretical studies on
SR for spike-train inputs
\cite{Chapeau96}-\cite{Hasegawa02a}.
The response of single IF neurons to coherent 
spike-train (ST) inputs 
is shown to be enhanced by an addition of
weak ST noises characterized by
the Poisson \cite{Chapeau96} \cite{Godivier96} or 
gamma distribution \cite{Mato98}.
Quite recently, the present author 
has studied SR of ensemble HH neurons
for {\it transient} spike-train inputs 
with independent Gaussian noise
by using the wavelet analysis \cite{Hasegawa02a}.

As for the third factor of (c) noises,
it has been reported that noises are ubiquitous in neural systems.
The origin of these noises is not clear at the moment. 
We may suppose, however, 
several conceivable origins of noises:
(i) cells in sensory neurons are exposed to
noises arising from the outer world,
(ii) ion channels of the membrane of neurons are known
to be stochastic \cite{Destexhe98},
(iii) the synaptic transmission yields noises originating 
from random fluctuations of the synaptic vesicle 
release rate \cite{Smith98}, and 
(iv) synaptic inputs include leaked currents from
neighboring neurons \cite{Shadlen94}.
Most of existing studies on
SR have simulated noises of the items (i)-(iii) by the Gaussian 
white noise \cite{Bulsara96}-\cite{Shimokawa99a}
\cite{Shimokawa99b}-\cite{Liu00}\cite{Hasegawa02a}
or Orstein-Uhrenbeck (OU) noise 
\cite{Longtin93}\cite{Lee98}\cite{Lee99}
\cite{Wang00}\cite{Liu00}.
ST noise is employed in Refs.\cite{Chapeau96}-\cite{Mato99}
taking account the item (iv).
In our study, we will include
the ST and white noises
which may be regarded, 
in a crude sense, as analog and digital noises, respectively,
with the rather different character.

One of the controversial issues concerning cortical neurons is
how the neurons may communicate information by spikes with 
a large variability. 
It has been reported that the variability of $c_v=0.5 \sim 1.0$
is observed in spike trains of non-bursting cortical neurons
in visual V1 and MT of monkey \cite{Softky92}, which is in strong contrast
with a small $c_v \;(= 0.05 \sim 0.1)$ in motor neurons \cite{Calvin68}. 
There have been much discussions how to understand 
the observed large variability:
a balance between excitatory and
inhibitory inputs \cite{Shadlen94},
the high physiological gain in the plot of 
input current vs. output frequency \cite{Troyer98}, 
correlation fluctuations in recurrent 
networks \cite{Usher94},
the active dendrite conductance \cite{Softky95},
input ISIs with the distribution of a slow-decreasing tail \cite{Feng98},
and input ISIs with large $c_v$ \cite{Hasegawa00}\cite{Brown99}.
We expect that although there might be several origins
responsible to the observed, large variability in ISI,
noises may be one of conceivable mechanisms.

%We expect that 
%noises might play some important roles in yielding
%a large variability.
%As will be shortly shwown in our simulations,
%ensemble neuron evoke variable firings
%when noises are added to sub-threshold input signals.
%Nevertheless SNR for this transmission is 
%fairly good because of SR effect.

The purpose of the present study is to investigate responses
of ensemble neurons to {\it spike-train} inputs subject to
ST and white noises, in order to get some insight to the 
following issues.

\noindent
(1) Is the effect of ST noise on responses to spike-train
signals, particalrly on SR, different from or same as 
that of white noise?

\noindent
(2) Is a population or ensemble of neurons 
important for the fidelity of signal transmission?

\noindent
(3) Is a large variability of spikes compatible with 
information transmission with a fairly good SNR? 

\noindent
Responses of single HH neurons to various types of spike-train
inputs with deterministic, chaotic and stochastic ISIs
without noises, have been investigated \cite{Hasegawa00}.  
SR for coherent spike-train inputs has been
theoretically studied
with single IF neurons \cite{Chapeau96}-\cite{Mato99}.
We should note, however, that 
the response to applied, external stimulus of 
the realistic HH model \cite{Hodgkin52}
is rather different
even qualitatively from that of the IF neuron 
\cite{Hasegawa00}\cite{Brown99}.
For an excitatory dc input $I$, the IF neuron 
which is classified as the type I, evokes the self-excited
oscillation showing the continuous $f_{o}-I$ relation
with a wide range of frequency $f_{o}$.
On the other hand, the HH neuron, which is classified as
the type II, has the discontinuous $f_{o}-I$ relation
at the critical current, above which it shows the
oscillation with a fairy narrow range of $f_{o}$.
For an inhibitory dc input current,
the HH neuron can fire with the so-called rebound process while
the IF neuron cannot.
Since the threshold-crossing behavior of the neuron is 
important in determining the behavior of its SR, 
it is necessary to re-examine SR
for the spike-train inputs with the use of the realistic
HH neuron model.
Furthermore, since SR of single neurons is generally different from that
of ensemble neurons, it is also necessary to
investigate SR not only of single HH neurons 
but also of ensemble HH neurons.

The present paper is organized as follows.
In Sec. II, an adopted model
for an ensemble of $N$-unit ($N$ = 1, 10, 100)
HH neurons is described.
Simulations for
responses of ensemble neurons to ST signals
with added ST noise, white noise, and ST plus white noises are
reported in Sec. IIIA, IIIB and IIIC, respectively,
where SR and the variability of ISIs are discussed. 
The final Sec. IV is devoted to conclusion and discussion.

%\newpage
\section{Ensemble Neuron Model}
%\begin{center}
%{\bf II. CALCULATION METHOD}
%\end{center}

%In order to study the propagation of spike trains in 
%neural networks, 
%we adopt a simple model of two layers
%which are referred to as the layer A and B, and
%which include $N_{A}$ and $N_{B}$ HH neurons, respectively.
%The HH neurons in the layer A receive a common
%input consisting of $M$ impulses
%and independent Gaussian noises.
%Output impulses of neurons in the layer A
%are feedforwarded 
%to neurons in the layer A through synaptic 
%couplings with time delays. 
%Our model mimics a part of the Sinfire nural network 
%with feedforward couplings but no intralayer ones.

We assume 
a network consisting of
$N$-unit HH neurons which receive
the same ST signals 
but independent, ST and Gaussian 
noises through excitatory synapses.
Spikes emitted by the ensemble neurons are collected
by a summing neuron.
A similar model was previously adopted by several
authors studying SR \cite{Collins95a}\cite{Pei96a}
\cite{Tanabe99}\cite{Hasegawa02a}.
Dynamics of the membrane potential $V_{i}$ of
the HH neuron {\it i}
is described by the non-linear differential
equations given by 

\begin{equation}
\bar{C} \:d V_{i}(t)/d t = -I_{i}^{\rm ion}  
+ I_{i}^{\rm ps} + I_{i}^{\rm n},
\;\;\;\;\;\;\;\;\;\;\;\;\;\; 
\mbox{(for $1 \leq i \leq N$)} \\
\end{equation}
where 
$\bar{C} = 1 \; \mu {\rm F/cm}^2$ is the capacity of the membrane.
The first term $I_{i}^{\rm ion}$ of Eq.(1) denotes the ion current
given by
\begin{equation}
I_{i}^{\rm ion} 
= g_{\rm Na} m_{i}^3 h_{i} (V_{i} - V_{\rm Na})
+ g_{\rm K} n_{i}^4 (V_{i} - V_{\rm K}) 
+ g_{\rm L} (V_{i} - V_{\rm L}),
\end{equation}
where
the maximum values of conductivities 
of Na and K channels and leakage are
$g_{\rm Na} = 120 \; {\rm mS/cm}^2$, 
$g_{\rm K} = 36 \; {\rm mS/cm}^2$ and
$g_{\rm L} = 0.3 \; {\rm mS/cm}^2$, respectively; 
the respective reversal potentials are   
$V_{\rm Na} = 50$ mV, $V_{\rm K} = -77$ mV and 
$V_{\rm L} = -54.5 $ mV.
Dynamics of the gating variables of Na and
K channels, $m_{i}, h_{i}$ and $n_{i}$,
are described by the ordinary differential equations,
whose details have been given elsewhere 
\cite{Hodgkin52}\cite{Hasegawa00}.

The second term $I_{i}^{\rm ps}$ in Eq.(1) denotes 
the post-synaptic current given by
\begin{equation}
I_{i}^{\rm ps} = \; \sum_{m}
g_{s}\:(V_a - V_s) \:\alpha(t-t_{{\rm i}m}), 
\end{equation}
which is induced by an input spike with the magnitude $V_{a}$ given by
\begin{equation}
U_{i}(t) = V_{a} \;\sum_{m}\; \delta(t-t_{{\rm i}m}),
\end{equation}
with the alpha function $\alpha(t)$:
\begin{equation}
\alpha(t) = (t/\tau_{\rm s}) \; e^{-t/\tau_{\rm s}} \:  \Theta(t).
\end{equation}
In Eqs.(3)-(5) $t_{{\rm i}m}= (m-1)\; T_{\rm s}$ is 
the $m$-th firing time with
the input-signal ISI of $T_{\rm s}$,
the Heaviside function is defined by
$\Theta (t)=1$ for $x \geq 0$ and 0 for $x < 0$,
%given by $t_{im} = (m-1) \: T_s$ for $m = 1 -M$, 
and $g_{s}$, $V_{\rm s}$ and $\tau_{\rm s}$ stand 
for the conductance, reversal potential and time constant,
respectively, of the synapse.

The third term $I_{i}^{\rm n}$ in Eq.(1) denotes  
added, independent noises which consist of two terms:
\begin{equation}
I_{i}^{n}(t) =  \sum_{m}\; \sqrt{C}\; \alpha(t-t_{im}^{n}) 
+ \sqrt{2 D}\;\xi_{i}(t).
%= \sqrt{2 D}\;\xi_{i}(t)
%+ \sum_{\ell}\;g_{n}(V_a-V_s)\;\alpha(t-t_{i\ell}^{n}).
\end{equation}
The first term of Eq.(6) expresses Poisson ST noise,
whose magnitude,
$C$, is hereafter
expressed by $\sqrt{C} \equiv g_{n}(V_a-V_s)$ in terms of $g_{n}$
as in Eq.(3) for a later purpose,
and $t_{im}^{n}$ is
the $m$-th firing time of the  ST noise
of the neuron $i$ with the average ISI of $\mu_{n}$.
The second term of Eq.(6) denotes 
Gaussian white noises with the magnitude of $D$
given by
\begin{equation}
< \overline{\xi_j(t)}> = 0, 
%< \overline{I_{i}^{n}(t)}> = 0, 
\end{equation}
\begin{equation}
<\overline{\xi_j(t) \:\xi_k(t')}> = \:\delta_{jk} \:\delta(t-t'),
%<\overline{I_{i}^{n}(t) \:I_{\ell}^{n}(t')}> 
%= 2 D \:\delta_{i\ell} \:\delta(t-t'),
\end{equation}
where the overline $\overline{ X }$ and 
the bracket $ < X >$ denote the temporal and
spatial averages, respectively. 
%The second term of Eq.(6) expresses
%spike-train noises 
%whose magnitude,
%$A$, is hereafter
%expressed by $A \equiv g_{n}(V_a-V_s)$ in terms of $g_{n}$
%as in Eq.(3) for a later purpose,
%and $t_{i\ell}^{n}$ is
%the firing time of the $\ell$-th spike-train noise
%of the neuron $i$, which is recurrently given by
%\begin{equation}
%t_{i \ell+1}^{n} = t_{i\ell}^{n} 
%+ T_{i\ell}^{n}(t_{i\ell}^{n}).
%\end{equation}
%We assume that the ISIs of spike-train
%noises $T_{ik}^{n}$ obey
%the gamma distribution:
%\begin{equation}
%P(T) = s^r\;T^{r-1}\; e^{-sT} /\;\Gamma(r),
%\end{equation}
%where $\Gamma(r)$ is the gamma function.
%The average and RMS of ISIs and 
%the variability are given by
%$\mu_{n}=r/s$, $\sigma_{n}=\sqrt{r}/s$
%and $c_{vn}=\sigma_{n}/\mu_{n} = 1/\sqrt{r}$, respectively.
%The initial firing times of $t_{i\ell}^{n}$
%with $\ell=1$ in Eq.(9) are assumed to be  
%randomly distributed in the range of $(0,\;\mu_{n})$.
%When $r=1$ in Eq.(9), 
%we have an exponential distribution
%for ISIs ($c_{vn}=1$) and a poisson distribution for
%the number of spikes in a given interval.
%In the limit
%of $r \rightarrow \infty$ and $s \rightarrow \infty$
%with keeping $\mu_n=r/s$, Eq.(9) reduces to
%$P(T)=\delta(T-\mu_n)$, periodic spikes with 
%$\mu_n=T$ and $c_{vn}=0$.

%The output spike of the neuron $i$ in an ensemble is given by
%\begin{equation}
%U_{{\rm o}i}(t) = V_{a} \;\sum_{n}\; \delta(t-t_{{\rm o}in}),
%\end{equation}
%in a similar form as an input spike [Eq.(4)],
%where $t_{{\rm o}in}$ is the $n$-th firing time when
%$V_{i}(t)$ crosses $V_{z}=0$ mV from below.

We should remark that our ensemble neuron model given 
by Eqs.(1)-(6) does not include synaptic couplings among
constituent HH neurons,
in contrast with the coupled ensemble models 
\cite{Kanamaru01}\cite{Wang00}\cite{Liu00};
related discussions being given in Sec. IV.
%As will be shown shortly, our ensemble neurons
%may show the enhancement in SNR 
%by the so-called {\it pooling effect} without couplings
%\cite{deCharms00}.

We assume that information is carried
by firing times of spikes.
Dividing the time scale by the width of time bin
of $T_{\rm b}$ as $t = t_{\ell}= (\ell-1) \;T_{\rm b}$ ($\ell$: integer),  
we define input and output
signals summed over ensemble neurons within the each time bin by
\begin{equation}
W_{\rm i}(t) = \sum_{m} \;\Theta(T_{\rm b}/2 - \mid t - t_{{\rm i}m}\mid),
\end{equation}
\begin{equation}
W_{\rm o}(t) = (1/N) \:\sum_{i=1}^{N} \sum_n \;
\Theta(T_{\rm b}/2 - \mid t - t_{{\rm o}in}\mid).
%\;\;\;\; \mbox{($1 \leq i \leq N$)}
\end{equation}
%\begin{equation}
%W_{\rm s}(t) = \sum_n \;
%\Theta(\mid t - t_{oin}\mid - T_b/2).
%\;\;\;\; \mbox{($i = N+1$)}
%\end{equation}
In Eqs. (9) and (10) $\Theta(t)$ stands for 
the Heaviside function,
%$W_{\rm i}(t)$ the external input signal
%(without noises),
%$W_{\rm o}(t)$ the output signal averaged over the ensemble neurons,
$t_{{\rm i}m}$ the $m$-th firing time of inputs, 
and $t_{{\rm o}in}$ the $n$-th
firing time of outputs of the neuron $i$ when
$V_{i}(t)$ crosses $V_{z}=0$ mV from below.
The time bin is chosen as $T_{\rm b}$ =2.5 ms in our simulations.
The fast Fourier transformation (FT) is performed for
$W_{\rm o}(t)$ in order to get
the SNR defined by
\begin{equation}
SNR = 10 \;{\rm log}_{10} (A_{s}/A_{n})\;\;\;\;\; \mbox{(dB)},
\end{equation}
where $A_{s}$ is the signal power spectrum at a given frequency
of $1/T_{\rm s}$ and $A_{n}$ the background noise level.

%We should remark that our model given by Eqs.(1)-(10) does 
%not include couplings among ensemble neurons.
%This is in contrast with some works on ensemble neurons
%\cite{Kanamaru01}\cite{Stocks01}\cite{Wang00}\cite{Liu00}
%where introduced couplings among 
%neurons play an important role in SR besides noises,
%related discussion being given in Sec. IV. 

Differential equations given by Eqs.(1)-(6)
are solved by the forth-order Runge-Kutta method
by the integration time step of 0.01 ms
with double precision.
The initial conditions for the variables are given by
%\begin{equation}
$V_{i}(t)$= -65 mV, $m_{i}(t)$=0.0526,  
$h_{i}(t)$=0.600, 
$n_{i}(t)$=0.313
at $t=0$,
%\end{equation}
which are the rest-state solution of a single
HH neuron.
Hereafter time, voltage, conductance,
current, and $D$
are expressed in units of ms, mV, ${\rm mS/cm}^2$, $\mu {\rm A/cm}^2$,
and $\mu {\rm A}^2/cm^4$, respectively.
We have adopted parameters of
$V_a=30$, $V_c=-50$,
and $\tau_{\rm s}$ =2.
%= $\tau_{n}=2$. 
%and $\tau_{i \ell}=10$.
Adopted values of $g_{s}$, $g_{n}$,
$D$, $\mu_{\rm n}$ and $N$ will be described shortly.
The simulation time for each run for a given set of parameters is 
$T_{sim} \sim$ 1500 ms (150000 $\times N$ time steps) 
and initial 3000 $N$ time steps are discarded to get asymptotic solutions.
The size of sample data for FT analysis becomes
$N_{FT}$=512 when the input ISI is chosen to be
$T_{\rm s}$ = 25 ms.
%for a confirmation of the accuracy. We incorporate
%OU noises given by Eqs.(8)-(10)
%using the method
%of Fox, Gatland, Roy and Vemuri \cite{Fox88}.
%The initial function for $V_{i}(t)$,
%whose setting is indispensable for
%the delay-differential equation, is given by
%\begin{equation}
%V_{j}(t)= -65 \:\: \mbox{\rm mV} 
%\:\: \mbox{\rm for} \:\: j=1,2
%\:\:  \mbox{at} \: t \in [-\tau_d, 0).
%\end{equation}
A single simulation with $N=100$
requires the CPU time of about 150 minutes by DOS/V PC
with 900 MHz processor.
The calculated SNR is expected to be improved 
if the simulation time $T_{sim}$ is increased.  
Unfortunately, we have not been able
to adopt larger $T_{sim}$ because of a limitation in
our computer facility (related discussion will be presented in Sec. IV). 
We hope that our simulation reveals some of the interesting
features of SR of ensemble neurons
for spike-train inputs with added, multiple noises.

%\newpage
\section{CALCULATED RESULTS}

%\vspace{0.5cm}
%\begin{center}
%{\bf III. CALCULATED RESULTS}
%\end{center}

\subsection{Spike-Train Noises}

\subsubsection{SNR and SR}

Firstly we discuss the case in which
ensemble HH neurons receive input
signals with independent ST noise only.
The input ISI is assume to be $T_{\rm s}$ = 25 ms
because spikes with this value of ISI are reported to be ubiquitous
in cortical brains \cite{Traub99}.
%The input synaptic strength is assume to be the same for all neurons:
%$g_i = g_{s}$. 
%When input synaptic strength is small:
%$g_s < g_{th}$, no neurons fire in the noise-free case,
%while it is sufficiently large ($g_s \geq g_{th}$) neurons fire,
%where $g_{th}=0.085$ is the threshold value.
We study sub-threshold inputs with $g_{s} < g_{th}$
for which neurons cannot fire 
without added noises.
The threshold value of $g_{th}$ generally depends 
on the input ISI, $T_s$:
for example, $g_{th}$=0.088, 0.085 and 0.095 
for $T_s$=20, 25 and 30 ms, respectively \cite{Note1}.

Raster in Fig. 1 shows firings of $N=100$ ensemble neurons
when spike-train signals of $T_{s}=25$ ms are applied to ensemble neurons
with added ST noises of $g_{n}=0.10$, $D=0.0$ and $\mu_{n}=25$ ms.
Neurons fire when a spike-train input plus noise 
or noises exceed the
threshold level.  At a glance, firings in Fig. 1 seem random.
We may realize it is not true when firings are summed over ensemble
neurons as shown in Fig. 2(b) where the output signal $W_{o}(t)$
summed over a $N=100$ ensemble is plotted. 
We note that $W_{o}(t)$ includes a periodic component
with a period of $T_{s}$=25 ms
as an input signal shown in Fig. 2(a).
This regularity is more clearly seen in its FT power spectrum 
shown in Fig. 3(a),
which shows clear peaks at a fundamental frequency of $1/T_{s}$ = 40 Hz
and its harmonics.
Information may be transmitted with aid of SR 
in a $N=100$ neuron ensemble.

When the size of ensemble neurons is small, however, the
information transmission is much degraded.
Figures 2(c) and 2(d) show $W_{o}(t)$ for $N=10$ and $N=1$,
respectively, other parameters except $N$ being the same as in Fig. 2(b).
Figure 2(d) for $N=1$ , for example, shows 
that a single neuron intermittently fires, yielding 18
firings for 40 input spike inputs.  
The FT power spectra for $N=10$ and $N=1$ are expressed in Fig. 3(b) and 3(c),
respectively.  
Figures 3(a)-3(c) show that the magnitude of a peak at the fundamental
frequency divided by that of background noises, {\it i.e.} SNR,
increases as the size of ensemble neurons is increased.
These mean that despite of ostensibly irregular firings of single neurons,
output signals summed over an ensemble, $W_{o}(t)$, become more regular
when $N$ becomes sufficiently large.

Figure 4 shows SNR as a function of $g_{n}$ 
calculated for $g_s=0.06$, $D=0$ and $\mu_{n}=25$ ms
with $N$ = 1, 10 and 100.
Results for $N$= 1and 10 are averages of ten runs.
When the noise intensity $g_n$ is increased from zero,
neurons begin to occasionally fire by a cooperative 
action between input signals and added noises.
As $g_n$ becomes greater than the threshold level of $g_{th}=0.085$,
ST noise alone is sufficient to trigger firings
without signal inputs.
When $g_n$ is further increased, SNR of outputs is gradually
degraded.  Then SNR has a maximum at $g_n \sim g_{th}$,
which is the characteristics of SR.
For the case of $N=10$ and 100, we realize SR with the maximum in SNR
for a weak noise of $g_{n} \sim 0.07$.
On the contrary, for the case of $N=1$, the maximum in SNR is not evident
although SNR is enhanced at $g_{n} > 0.04$. 

In many non-linear systems, SNR for sinusoidal input signals
is reported to obey the noise-intensity dependence 
given by \cite{Gammai98}\cite{Anish99}
\begin{equation}
SNR = 10 \; {\rm log}_{10}\;[(A/X^{\alpha}) \; {\rm exp}(-B/X)],
\end{equation}
with a maximum at $X_{max}=B/\alpha$ where
$X=D$ (the intensity of the white noise), $\alpha=2$, 
and A and B are constants depending on the model parameters.
Substituting $X=g_{n}^2$ [$\propto C$ in Eq.(6)], we have tried to analyze
the $g_{n}$ dependence of SNR for $N=100$
obtained in our simulation.
Dashed curves in Fig.4 show the results adopting
two sets of parameters in Eq.(12):
($\alpha, A, B$)=(2, $1.77 \times 10^{-2}$, $8.45 \times 10^{-3}$) 
and  (3, $2.00 \times 10^{-4}$, $1.27 \times 10^{-2}$),
which are chosen such as to locate the maximum of SNR
at $g_n=0.065$.
The latter choice of parameters with $\alpha=3$
yields the much better agreement
with the data obtained in our simulations than the former
with $\alpha=2$, although the plateau around the maximum
is not well reproduced even in the latter.

We have investigated the effect of input ISI on SR 
by changing $T_{s}$. 
Solid curves in Fig. 5 denote SNRs as a function of $g_{n}$
for $T_{s}$=20, 25 and 30 ms
obtained in our simulation and
dashed curves those analyzed by Eq.(12) with $X=g_{n}^2$ and
sets of parameters of 
($\alpha, A, B$)
=(3, $5.01 \;\times 10^{-4}$, $1.27 \times 10^{-2}$) for $T_{s}=20$,
(3, $2.00 \times 10^{-4}$, $1.27 \times 10^{-2}$) for $T_{s}=25$, and
(3, $7.94 \;\times 10^{-5}$, $1.27 \times 10^{-2}$) for $T_{s}=30$.
The index $\alpha=3$ is realized for all the $T_{s}$
values investigated.
The maximum SNR value is decreased as $T_{s}$ is increased, although
the maximum position at $g_{n} \sim 0.065$ in SNR is not changed.

%suprathreshold inputs

So far our discussion
is confined to the sub-threshold case.
We have performed simulations also for the supra-threshold inputs,
adopting $g_{s}$ larger than the threshold value of $g_{th}$.
Figure 6 shows 
the $g_s$ dependence of SNR 
calculated for ST noise only ($g_n=0.10$ and $D=0$)
with $\mu_{n}=25$ ms, 
and $N=1$, 10 and 100.
When $g_s$ is increased across the threshold value of $g_{th}$,
SNR is discontinuously increased.
SNR for the supra-threshold inputs is better than that for
sub-threshold inputs, as expected.
We note, however, that ensemble neurons with large $N$ is fairly
robust against weak noises relevant to SR.

\subsubsection{ISI}

Next we discuss the distribution of output ISIs of
{\it individual} neurons given by 
\begin{equation}
T_{oin}=t_{oin+1}-t_{oin},
\end{equation}
where $t_{oin}$ is the $n$-th firing time of outputs 
of the neuron $i$.
Figures 7(a)-(d) show histograms of output ISIs
for $g_{s}$=0.02, 0.06, 0.10 and 0.14,
respectively, when ST noise of $g_{n}=0.10$ is added to 
ST signal (see Fig. 6).
In the cases of $g_{s}=0.02$, ISI histograms (ISIH) nearly
obey the exponential distribution as shown by
dashed curve, but it vanishes
at $T_{oin} < 15$ ms because 
a HH neuron cannot properly respond to small-ISI inputs
due to its refractory period \cite{Hasegawa00}.
On the contrary, in the cases of $g_s=0.10$ and 0.14,
ISIH has larger magnitudes at $T_{oin} \sim T_s$, 
which may be approximately 
expressed by the gamma-type distribution.
In the case of $g_{s}=0.06$, ISIH includes not only the truncated 
exponential distribution but also finite contributions
at multiples of $T_{s}$. 
This change in the distribution is more clearly seen in
the $g_{s}$ dependence of
the average ($\mu_o$) and RMS values ($\sigma_o$)
of output ISIs, which are plotted in Fig. 8.  
Both $\mu_o$ and $\sigma_o$ are almost constant at
$g_s < 0.08$, and at $g_s \simg 0.10$ they are suddenly decreased. 
When ISIH of output spikes for $g_{s} < g_{th}$ is 
expressed by the truncated
exponential distribution given by
\begin{equation}
P(T) \propto \Theta(T-T_{L}) \;{\rm exp}(-s T),
\end{equation}
$\Theta(\cdot)$ being the Heaviside function and $T_{L}$
the lower bound, the average and RMS values are
given by $\mu_{o}=(1+s T_{L})/s$ and $\sigma_{o}=1/s$,
which yield $\mu_{o}=40$ and $\sigma=25$ ms
for $1/s=25$ and $T_{L}=15$ ms.  These figures are 
a little different from
$\mu_{o} \sim 50$ and $\sigma \sim 35$ ms
for $g_{s} < 0.8$ shown in Fig. 8. 
The difference may be attributed to the extra contribution
at $2\;T_{s}$ obtained in our simulation, as shown in Figs. 7(a) and (b).
The variability defined by $c_{vo} = \sigma_{o}/\mu_{o}$ 
is 0.655, 0.703, 0.307 and 0.212 
for $g_{s}$=0.02, 0.06, 0.10 and 0.14, respectively.
When comparing Fig. 6 and Fig. 8, we note that even if
the variability of ISIs of individual neurons 
is considerable, output signals summed over
a large-scale ensemble [$W_{o}(t)$ in Eq.(10)] may carry information
with a fairly good SNR.  
For example, in the case of $g_{s}=0.06$,
we get SNR=17.4 dB for $c_{vo}$ = 0.703.
This is an advantage of a neuron ensemble.

\subsection{White Noises}

\subsubsection{SNR and SR}

Next we discuss SR for Gaussian white noise, which is 
applied to our ensemble neurons instead of ST noise.
Neurons occasionally fire when signal plus noise
exceed the threshold level. 
Firings of neurons for white noises are similar to those
to the ST noises shown in Fig. 1.
SNR calculated as a function of $D$, the intensity of the white noise, 
for $N=1$, 10 and 100 is plotted in Fig. 9.
When the white noise intensity is increased from zero,
SNR is rapidly enhanced with a maximum at $D \sim 2$
followed by a gradual decrease, which is a typical SR.
Although the calculated SNR shows the 
SR behavior irrespective of N, it is more evident for larger $N$.

The dashed curve in Fig. 9 expresses SNR calculated by Eq.(12)
with a set of parameters of ($\alpha, A, B$)=(2, 5890, 5), 
which are chosen such as to 
agree with the maximum position at $D=2.5$ in SNR for $N=100$
obtained by our simulation.   
The agreement between the result of $\alpha=2$
and our data seems satisfactory.
This value of $\alpha=2$ agrees with
the results of SR for sinusoidal input signals in HH neurons 
\cite{Lee99}\cite{Liu00}
as well as those realized in many non-linear systems
\cite{Gammai98}\cite{Anish99}.

%suprathreshhold inputs

Fig. 10 shows the $g_{s}$ dependence of SNR for white noises only
($D=2$ and $g_{n}=0$) with $N=1$, 10 and 100.
SNR is gradually increased as increasing $g_{s}$.
In contrast with the case for ST noise shown in Fig. 6,
there is no significant changes in SNR at the threshold
level shown by the vertical, dashed line.
A comparison of Fig. 6 and Fig. 10 shows that 
$g_{s}$ dependence of SNR for ST noise is different from 
that for white noise.

\subsubsection{ISI}

Figures 11(a)-(d) show histograms of output ISIs
for $g_{s}$=0.02, 0.06, 0.10 and 0.14,
respectively, when white noise of $D=2$ is added to 
ST signal (see Fig. 10).
ISIHs for $g_{s}=0.02$ and $g_{s}=0.06$
show a typical behavior with peaks
at multiples of $T_{s}$ whose magnitudes 
decrease exponentially \cite{Longtin93}\cite{Longtin91}.
It is noted that the distribution
extends up to 200 ms.
As $g_{s}$ is increased across $g_{th}$,
magnitudes of the main peak at $T_{s}$ are much increased as expected. 
Figure 12 shows the $g_{s}$ dependence of
the average ($\mu_{o}$) and RMS values ($\sigma_{o}$) of ISI
and the variability ($c_{vo}$).
As increasing $g_{s}$, both $\mu_{o}$ and $\sigma_{o}$ 
are gradually decreased with
no sudden changes at $g_{s} \sim g_{th}$.
The variability
is 0.813 and 0.710 for $g_{s}$=0.02 and 0.06, respectively.
As $g_{s}$ becomes larger than $g_{th}$, 
ISIH has a larger magnitude at $T_{oin} \sim T_{s}$ and
then the variability is decreased:
$c_{vo}$ becomes 0.528 and 0.343
for $g_{s}$=0.10 and 0.14, respectively.
%It is noted that in a large neuron ensemble,
%sub-threshold inputs may be transmitted
%with a fairly good SNR improved by SR, even though they yield
%a large variability in their output ISIs.
Comparing Fig. 12 to Fig. 8, we note that
$g_{s}$ dependence of $\mu_{o}$, $\sigma_{o}$ and $c_{vo}$
for white noises is rather different from that for ST noise.

\subsection{Spike-Train plus White Noises}

Since neurons in real neural systems are in the environment
with various kinds of noises as discussed in Sec. I,
it is necessary to examine various effects 
of multiple types of noises.
Taking into account independent Gaussian and OU noises,
Liu, Hu and Wang have investigated the effect of
spatial correlation on SR in coupled HH neurons \cite{Liu00}.
Recently Lindner and Schimansky-Geier\cite {Lindner01} have
included, in the ensemble IF model, the additive and signal-coded noises, 
which are expressed by $\sqrt{2 D_1} \;\xi_1(t)$ 
and $\sqrt{D_2\:s(t)} \;\xi_2(t)$,
respectively, in terms of the external sinusoidal signal $s(t)$ and 
Gaussian noises $\xi_n$ with the magnitudes of $D_n$ ($n=1,2$).

In previous Sec. IIIA and Sec. IIIB, 
we have separately discussed ST and white noises.
Now we simultaneously add both the noises to our ensemble neurons.
Figure 13 shows the three-dimensional plot of SNR for $N=100$ as functions of
$g_{n}$ and $D$.
In the case of $D=0$ (ST noise only),
SNR has a maximum at $g_{n} \sim 0.07$,
as shown in Fig.4.
In the case of $g_{n}=0$ (white noises only), on the other hand,
SNR has a maximum
at $D \sim 2$ as shown in Fig. 9.
The contour plot depicted in the base of Fig. 13 shows that SNR is rapidly
increased from zero as ST or white noise is increased.
SNR in the presence of weak white noise of $0 < D < 1$
is enhanced by a further addition of
ST noise and it depends considerably on $g_n$.
With stronger white noise of $D > 1$, 
SNR is slightly enhanced by an addition of 
weak ST noise of $g_{n} \sim 0.01$ although it only weakly
depends on $g_n$ at $g_{n} > 0.02$.
SNR in the presence of weak ST noise with $g_{n} < 0.05$
is much enhanced by a further addition of white noise.
We note, however, that white noise enhances SNR even 
in the presence of stronger ST noise of $g_{n} > 0.1$,
where SNR is decreased by excess ST noise for $D =0$. 
These clearly show that white noise is more effective
for SR than ST noise.

\section{CONCLUSION AND DISCUSSION}

%\vspace{0.5cm}
%\begin{center}
%{\bf VI. CONCLUSION AND DISCUSSION}
%\end{center}

In previous Sec. III, the simulation time of each run has
been limited to be 1500 ms (150000 $\times N$ time steps) 
because of a limitation
in our computer facility.  It is, however, possible, 
to extend the simulation time when the simulation
is made only for single ($N=1$) neurons.
Figure 14 shows SNR as a function of $g_n$ 
for ST noise added to single HH neurons,
which are calculated with
the simulations times of $T_{sim}=$ 1500, 3000, 6000
and 12 000 ms, yielding FT-data sizes ($N_{FT}$)
of 512, 1024, 2048 and 4096, respectively. 
We note that as $T_{sim}$ becomes larger, SNR is 
improved, in particular, its maximum becomes more evident. 
The dashed curve expresses SNR calculated by using Eq.(12)
with $X=g_{n}^2$ and a set of parameters of $(\alpha, A, B)$
=(3, $6.31 \times 10^{-5}$, $9.75 \times 10^{-3}$), 
which are chosen to reproduce
SNR calculated for $T_{sim}=12 000$ ms with the maximum at $g_{n}=0.057$.  
Again the index of $\alpha=3$ is realized for ST noise
added to single neurons. We note from
Figs. 4 and 14 that SR becomes more significant if the duration
of applied, coherent spike trains is longer and/or the size of
ensemble neurons is larger.
Even if the duration of applied signal is not long, SNR may
be improved if the size of ensemble neurons is sufficiently large.
This is more evident when the input signal is {\it transient}
spike train, as recently demonstrated in Ref.[54].  

In a summary, we have numerically investigated SR
responses of an ensemble of HH 
neurons to spike-train signals with added ST and/or
white noises. 
Our conclusions against the three issues raised in
the Introduction are summarized as follows:
 
\noindent
(1) Comparisons of Figs. 4, 6, 7 and 8 for ST noise
with Figs. 9, 10, 11 and 12 for white noise,
respectively, clearly show both the difference and 
the similarity between the effects of
ST and white noises. 
Although SR is a genetic phenomenon,
its detailed behavior depends on kinds of the input signal 
and added noises. 
When analyzing SNR obtained in our simulations with
the use of Eq.(12),
we get $\alpha=3$ for ST noise, which is different
from $\alpha=2$ for white noise.
ST noise is less effective for SR 
than the white noise (Fig. 13).

\noindent
(2) SNR is more improved as the size of ensemble is larger. 
In a large neuron ensemble,
the transmission
fidelity for supra-threshold inputs
is not significantly degraded by weak noises
responsible to SR for sub-threshold inputs (Figs. 6 and 10).

\noindent
(3) The variability of ISIs of individual neurons
for sub-threshold inputs is rather large ($c_{vo} \siml 0.8$).
Nevertheless the output $W_{o}$ summed over an ensemble
may carry information
with a fairly good SNR.

\noindent
The item (2) is consistent with the results of SR 
for transient spike-train signal \cite{Hasegawa02a} and
for analog signals \cite{Collins95a},
showing that a population
of neurons plays a very important role for the transmission
of spike-train inputs both with sub- and supra-threshold levels.
It is worth to note that this enhancement in SNR 
is due to the {\it pooling effect} \cite{deCharms00}
because our ensemble neuron model have no couplings
among HH neurons.
The item (3) shows that noise may be one of conceivable mechanism
yielding a large variability observed experimentally \cite{Softky92}.
Even when the variability of firings of individual neurons is considerable, 
firings summed over a ensemble may carry information with 
a fairly good SNR enhanced by SR and pooling effects.
Thus the large variability and high SNR are not incompatible
in a large-scale neuron ensemble.

The present study entirely relies on
simulations.
It would be interesting to theoretically
elucidate the dependence of the index $\alpha$
on a kind of added noises mentioned above. 
However,
conventional approaches having been employed for 
a study of SR such as the rate-equation and 
linear-response theories \cite{Gammai98}-\cite{Anish99},
do not work on our case.
%Mato \cite{Mato98} adopted Gammaitoni's approach \cite{Gammaitoni95}
%for an analysis of his SR result
%for continuous spike-train signals with ST noises.
%It seems, however, not to be transposed directly
%to our case
%even if our HH model is replaced by simpler IF model
%or threshold-crossing model.
We leave its analytical study as our future problem.

\section*{Acknowledgements}
This work is partly supported by
a Grant-in-Aid for Scientific Research from the Japanese 
Ministry of Education, Culture, Sports, Science and Technology.

%\begin{references}

\begin{figure}
\caption{
%Fig. A 
Raster showing firings in $N=100$ ensemble neurons
with $D=0.0$, $g_n=0.10$, $\mu_{n}=25$ ms
and $g_s=0.06$ (ST noise only).
}
\label{fig1}
\end{figure}

\begin{figure}
\caption{
%Fig. B
(a) An input signal $W_{i}(t)$, and
(b) output signals $W_o(t)$ for $N=100$, (c) $N=10$ and 
(d) $N=1$,
with $D=0.0$, $g_n=0.10$, $\mu=25$ ms 
and $g_s=0.06$,
results of (b) and (c) being multiplied by factors of 
5 and 2, respectively.
}
\label{fig2}
\end{figure}

\begin{figure}
\caption{
%Fig. C
Fourier power spectra for (a) $N=100$, (b) $N=10$ and (c) $N=1$,
with $D=0.0$, $g_n=0.10$, $\mu_{n}=25$ ms
and $g_s=0.06$.
}
\label{fig3}
\end{figure}

\begin{figure}
\caption{
%Fig. D. 
The $g_n$ dependence of SNR for ST noise
with $D=0.0$, $\mu_{n}=25$ ms
and $g_{s}=0.06$,
results for $N=1$ and 10 being averages of ten runs.
Dashed curves express SNR calculated by Eq.(12) with $X=g_{n}^2$
for $\alpha=2$ and 3 (see text).
}
\label{fig4}
\end{figure}

\begin{figure}
\caption{
%Fig. O. 
The $g_n$ dependence of SNR for ST noise and input signals with
$T_{s}=$20, 25 and 30 ms
($N=100$, $D=0.0$, $\mu_{n}=25$ ms and $g_{s}=0.06$),
dashed curves expressing SNR calculated by Eq.(12) with $X=g_{n}^2$
for $\alpha=$ 3 (see text).
}
\label{fig5}
\end{figure}

\begin{figure}
\caption{
%Fig. E. 
The $g_s$ dependence of SNR for ST noises
with $D=0.0$, $g_n=0.10$ and $\mu_{n}=25$ ms,
results for $N=1$ and 10 being averages of ten runs
and the dashed line the threshold value of
$g_{th}=0.085$.
}
\label{fig6}
\end{figure}

\begin{figure}
\caption{
%Fig. J. 
Histograms of output ISIs of $N=100$ ensemble neurons
for (a) $g_{s}=0.02$, (b) 0.06, (c) 0.10 and (d) 0.14,
with input signals of $T_{s}$=25 ms and
added ST noise ($g_{n}$=0.10, $D=0$).
Dashed curves in (a) and (b) express the exponential distribution
given by $P(T) \propto {\rm exp}(-T/25)$ and
histograms of (c) and (d) are multiplied by a factor of 1/5.
}
\label{fig7}
\end{figure}

\begin{figure}
\caption{
The $g_{s}$ dependence of the average ($\mu_{o}$)
and RMS values ($\sigma_{o}$)  of ISI, and 
the variability ($c_{vo}$) 
when ST noise of $g_{n}=0.10$ is added
to $N=100$ ensemble,
the dashed line denoting 
the threshold value of $g_{th}=0.085$.
}
\label{fig8}
\end{figure}

\begin{figure}
\caption{
%Fig. I. 
The $D$ dependence of SNR for white noises
with $g_n=0.0$, $\mu_{n}=25$ ms
and $g_{s}=0.06$, 
results for $N=1$ and 10 being averages of ten runs.
The dashed curve expresses SNR calculated by Eq.(12) with $X=D$ 
for $\alpha=2$ (see text).
}
\label{fig9}
\end{figure} 

\begin{figure}
\caption{
%Fig. J. 
The $g_s$ dependence of SNR for white noises
with $D=2.0$, $g_{n}=0.0$ and $\mu_{n}=25$ ms,
results for $N=1$ and 10 being averages of ten runs
and the dashed line the threshold value of
$g_{th}=0.085$.
}
\label{fig10}
\end{figure}

\begin{figure}
\caption{
%Fig. J. 
Histograms of output ISIs of $N=100$ ensemble neurons 
for (a) $g_{s}=0.02$, (b) 0.06, (c) 0.10 and (d) 0.14,
with input signals of $T_{s}$=25 ms and
added white noise ($D=2$, $g_{n}$=0),
histograms of (c) and (d) being multiplied by a factor of 1/3.
}
\label{fig11}
\end{figure}

\begin{figure}
\caption{
The $g_{s}$ dependence of the average ($\mu_{o}$)
and RMS values ($\sigma_{o}$) of ISI, and 
the variability ($c_{vo}$) 
when white noises of $D=2.0$ is added
to $N=100$ ensemble,
the dashed line denoting 
the threshold value of $g_{th}=0.085$.
}
\label{fig12}
\end{figure}

\begin{figure}
\caption{
%Fig. M. 
The three-dimensional plot of SNR
as functions of $g_{n}$ and $D$ for $\mu_{n}$=25 ms and $g_{s}=0.06$,
the contour plot being shown in the base of the figure.
}
\label{fig13}
\end{figure}

\begin{figure}
\caption{
%Fig. J. 
The $g_n$ dependence of SNR for single ($N=1$)
HH neurons with ST noise
calculated by changing $T_{sim}$ (ms),
the simulation time for each run 
($D=0$, $\mu_{n}=25$ ms and $g_s=0.06$),
the result for $T_{sim}=1500$ ms being the average of ten runs.
The dashed curve denotes SNR given by Eq.(12) 
with $X=g_{n}^2$ and $\alpha=3$ (see text).
}
\label{fig14}
\end{figure}

\end{document}